%% file: Manuscript.tex
\documentclass[fleqn,usenatbib]{mnras}

\usepackage{newtxtext,newtxmath}

\usepackage[T1]{fontenc}

\DeclareRobustCommand{\VAN}[3]{#2}
\let\VANthebibliography\thebibliography
\def\thebibliography{\DeclareRobustCommand{\VAN}[3]{##3}\VANthebibliography}

\input{Packages}

\input{Commands}

\title[
    Improved lanthanide constraints for AT\,2017gfo
]{
    Improved lanthanide constraints for the kilonova AT\,2017gfo
}

\author[
    Gillanders, Fl\"ors \& Ferreira~da~Silva
]{
    J.~H.~Gillanders\,\orcidlink{
        0000-0002-8094-6108
    }$^1$\thanks{
        E-mail: \href{mailto:james.gillanders@physics.ox.ac.uk}{james.gillanders@physics.ox.ac.uk}
    },
    A.~Fl\"ors\,\orcidlink{0000-0003-2024-2819}$^2$ and
    R.~Ferreira~da~Silva\,\orcidlink{0000-0003-3030-0496}$^{3, 4}$ \\
    $^1$Astrophysics sub-Department, Department of Physics, University of Oxford, Keble Road, Oxford, OX1 3RH, UK \\
    $^2$GSI Helmholtzzentrum f\"ur Schwerionenforschung, Planckstra{\ss}e~1, D-64291~Darmstadt, Germany \\
    $^3$Laboratório de Instrumentação e Física Experimental de Partículas (LIP), Av. Prof. Gama Pinto 2, 1649-003 Lisboa, Portugal \\
    $^4$Faculdade de Ciências da Universidade de Lisboa, Rua Ernesto de Vasconcelos, Edifício C8, 1749-016, Lisboa, Portugal
}

\date{Accepted XXX. Received YYY; in original form ZZZ}

\pubyear{\the\year{}}

\begin{document}
\label{firstpage}
\pagerange{\pageref{firstpage}--\pageref{lastpage}}
\maketitle

\begin{abstract}
    Spectroscopic observations of the kilonova AT\,2017gfo provide a unique opportunity to identify signatures from individual heavy elements freshly synthesised via the {\it r}-process, the nucleosynthetic channel responsible for producing $\sim$\,half of all trans-iron-group elements. Limitations in the available atomic data have historically hampered comprehensive line identification studies; however, renewed interest has led to the generation of improved (more complete and accurately calibrated) line lists for {\it r}-process species. Here we demonstrate the utility of such data, by exploiting newly generated line lists for the lanthanides to model the photospheric-phase 3.4\,d X-shooter spectrum of AT\,2017gfo with the radiative transfer tool \textsc{tardis}. We find the data can only be reproduced by invoking a substantially diminished lanthanide mass fraction ($X_{\textsc{ln}}$) than that proposed by previous studies. Specifically, our model necessitates $X_{\textsc{ln}} \approx 2.5 \times 10^{-3}$ in the line-forming region, a value $20 \times$ lower than previously claimed. This substantial reduction in $X_{\textsc{ln}}$ is driven by our inclusion of much more complete lanthanide line information that enables better estimation of their total contribution to the observations. We encourage future modelling works to exploit all atomic data advances, and also encourage continued efforts to generate the necessary data for the remaining {\it r}-process species of interest.
\end{abstract}



\section{Introduction}

For more than fifty years, compact binary mergers have been postulated as an ideal site for the rapid neutron-capture process (\rpro) to synthesise many of the heaviest elements \citep{Lattimer1974, Symbalisty1982, Eichler1989, Li1998, Freiburghaus1999, Rosswog1999}. Simulations of a neutron star (NS) merging with another NS, or a stellar-mass black hole (BH), have demonstrated that the conditions within the ejected material can plausibly support this extreme nucleosynthetic channel \citep[\eg,][]{Goriely2011, Goriely2013, Goriely2015, Korobkin2012, Perego2014, Wanajo2014, Just2015}. The radioactive decay of the ensemble of freshly synthesised unstable heavy elements produced via \rpro\ nucleosynthesis can subsequently power a thermal ultraviolet--optical--infrared transient, dubbed a kilonova \citep[KN;][]{Li1998, Metzger2010, Barnes2013, Metzger2019}.

Since the observations of the first spectroscopically confirmed kilonova event \citep[\gfo;][]{MMApaper2017}, numerous works have attempted to model these data in an effort to identify spectral features that can be linked to an individual (or a family of) \rpro\ species, leading to direct tethering of \rpro\ production to neutron star mergers. Initial work by \cite{Smartt2017} suggested Cs and Te were dominant contributors to the observed $0.7 - 0.85$\,\micron\ absorption feature in the early spectra of \gfo. Further work by \cite{Watson2019} showed that this absorption was likely produced instead by the first \rpro\ peak element Sr, and not Cs or Te. This Sr identification has since been independently corroborated by further quantitative analyses \citep[\eg,][]{Domoto2021, PaperI, Perego2022}. Further spectral identifications have been proposed for this (and other) features in the spectra; some examples include Y \citep[][]{Sneppen2023_YII}, La and Ce \citep[][]{Domoto2022, PaperII}, Te \citep[][]{Hotokezaka2023, PaperII}, Rb \citep[][]{Pognan2023} and He \citep{Perego2022, Tarumi2023}. All feature identification studies have suffered from limitations in the available line transition data, with works typically focusing on the species that have reasonably complete data, or exploiting newly generated data.

The interpretation of kilonova spectra relies on atomic data that are both complete and accurate. These two qualities, however, do not always come from the same source. Theoretical atomic structure calculations are indispensable because they can, in principle, provide line lists that are essentially complete for a given ion \citep[\eg,][]{Tanaka2020}. This completeness is crucial for opacity calculations; if important transitions are missing, the total opacity of the ejecta will be underestimated, resulting in misleading conclusions about the physical conditions and the elemental composition of the outflow. By contrast, experimental measurements of energy levels and transition wavelengths are limited in scope due to the wavelength range that can be sampled, but also by the sheer numbers of dense lines in the experimental spectra of open $f$-shell species that can lead to multi-year line identification studies for a single ion. As a result, these open $f$-shell species typically were not (until recently) studied in detail. Oscillator strengths are harder still, since they require radiative lifetime measurements combined with branching fractions, or calibrated absorption experiments. Thus, relying exclusively on the existing experimental data can severely underestimate the true opacity.

Theoretical calculations face their own limitations. While calculations optimised for generating large data sets provide completeness, the predicted wavelengths of transitions are often inaccurate. In extreme cases, uncalibrated calculations can deviate from laboratory measurements by as much as 50~per~cent \citep[see \eg,][]{Flors2023, Shingles2023}. Even much smaller discrepancies are enough to make the reliable identification of individual spectral features very challenging. Consequently, data sets based solely on theoretical calculations can be used to compute opacities, but are inadequate for detailed line identification studies.

Taken together, these issues highlight the need for a combined approach. For reliable modelling of kilonova spectra, one requires the completeness of theoretical atomic structure calculations to capture the total opacity, alongside the accuracy of experimental measurements to allow confident line identifications for spectral features. Calibrating theoretical data against available experimental benchmarks is therefore essential, since it bridges the gap between completeness and accuracy. Only through this synthesis can we achieve the dual goals of computing reliable opacities, and securely identifying the species responsible for the spectral feature(s) present in KNe.

Significant advances have been made in expanding the available atomic data needed for detailed KN studies. A recent example includes the publication of the \textit{Japan-Lithuania Opacity Database for Kilonova}\footnote{\url{http://dpc.nifs.ac.jp/DB/Opacity-Database}} \citep[][]{Tanaka2020, Kato2024}, which contains uncalibrated line transition information for neutral to triply ionised Fe to Ra ($Z = 26 - 88$, ions \I--\IV). These data are well-suited for capturing the general impact of these heavy elements to the opacity of KN ejecta, but transition wavelength uncertainties prevent precise spectral feature predictions. Other atomic physics groups have also been working towards generating the line transition information needed for \rpro\ species; for example, calibrated line information now exists for some of the first \citep[Sr, Y and Zr;][]{Mulholland2024_Sr_Y, Dougan2025_Sr, McCann2025_Zr}, second \citep[Te;][]{Mulholland2024_Te, Mulholland2025_TeIV+V_arXiv} and third \citep[W, Pt and Au;][]{Smyth2018_WI, Gillanders2021, Dunleavy2022_WII, McCann2022, McCann2024_WIII} \rpro\ peak elements.

Recently, \cite{Flors2026} presented calibrated line information for singly and doubly ionised lanthanide species ($_{57}$La -- $_{70}$Yb). This catalogue of atomic structure data comprises 146,856 bound energy levels and nearly 28.7~million transitions across 28 ions, all computed with the \texttt{Flexible Atomic Code} \citep[\texttt{FAC};][]{Gu2008_FAC}. Of these, 66,591 transitions were experimentally calibrated and therefore provide highly reliable wavelengths. The strong electric dipole transitions with $\log(gf) > -1$ show good agreement with available experimental and semi-empirical data. Moreover, for ions with relatively extensive experimental information, the resulting opacities are consistent with previous models, while also capturing the wavelength accuracy of low-lying transitions where experimental data are available.

In this manuscript, we explore the effects of extending previous radiative transfer studies to make use of recent advances in atomic data. Specifically, this study aims to explore the inferences one can now make by utilising the current state-of-the-art, publicly available atomic data presented by \cite{Flors2026}. To this end, we re-analyse the 3.4~day \xsh\ spectrum of \gfo\ in a similar manner to that of \cite{PaperI}. We present similar models and focus specifically on the differences caused by incorporating new, more complete, line transition information.

\section{Atomic data} \label{SEC: Atomic data}

For our radiative transfer modelling we utilise \tardis, which we describe in Section~\ref{SEC: Model setup}. First, we outline the different atomic data sources that we combine for our modelling efforts.

In addition to the sources used to build the `default' \tardis\ atomic data file \citep[the \chianti\ and Kurucz \kuruczgfall\ data;][]{Chianti-OG, Chianti-v9, Kurucz1995_CD23}, we expand to include the Kurucz \kuruczatoms\ \citep[][]{Kurucz2018}, \dream\ \citep[][]{DREAM1, DREAM2}, \qub\ \citep[][]{Gillanders2021, McCann2022}, and \gsi\ atomic data \citep{Flors2026}. We substitute the existing (or non-existent) `default' \tardis\ atomic data following a hierarchical preference of \mbox{\gsi\ {\scriptsize \&} \qub\ $\rightarrow$ \dream\ {\scriptsize \&} \kuruczatoms$_{\rm Sr - Zr}$ {\scriptsize \&} \chianti$_{\rm H, \, He}$ $\rightarrow$ \kuruczgfall}. A summary of which atomic data sources are included for $Z = 1 - 92$ is presented in Table~\ref{TAB: Atomic data}. We note that the atomic data compilation presented here is almost identical to that utilised by \cite{PaperI}; the only difference is that here we have replaced the \dream\ {\scriptsize \&} \kuruczgfall\ data for $Z = 57 - 70$ (ions \II\ and \III) with that of \gsi, which represents an improvement in both completeness and accuracy \citep[see][]{Flors2026}.

These atomic data sources were combined and ingested into an atomic data file for use with \tardis\ using the \carsus\ tool.\footnote{\url{https://github.com/tardis-sn/carsus}} We truncate all line lists by discarding transitions with ${\rm log}(gf) < -3$ to reduce computational demand. These weak transitions have a negligible impact on our synthetic spectra, and so they can be safely discarded.\footnote{We ran test models with an extended line list including lines for \mbox{${\rm log}(gf) \geq -5$}, and found that these additional lines had a negligible impact on the resultant synthetic spectrum.}

After compiling our new atomic data set, we find that the total number of lines included has substantially increased relative to the atomic data set utilised by \cite{PaperI}. Given the only difference in these data compilations is the inclusion of \gsi\ data over \dream, we compare the number of lines in each (see Table~\ref{TAB: GSI_vs_DREAM}). The \gsi\ data set contains 17,897,242 lines for singly and doubly ionised $_{57}$La -- $_{70}$Yb, a $370 \times$ increase in the number of lines contained within \dream\ for the same species and for the same $\log(gf)$ cut (48,633).

\input{Tables/AtomicDataSources}

\input{Tables/GSI_vs_DREAM}

\section{Model setup} \label{SEC: Model setup}

As in \cite{PaperI}, we use \tardis\ \citep[][]{Kerzendorf2014_TARDIS}, the one-dimensional, time-independent, Monte Carlo radiative transfer spectral synthesis tool for generating our synthetic spectra.\footnote{Specifically, we use \tardis~\texttt{v2024.01.08}.} While \tardis\ was originally developed to model thermonuclear supernovae (SNe), the code has since been exploited to model a range of explosive transient phenomena, with multiple works utilising the tool to generate synthetic KN spectra \citep[\eg,][]{Smartt2017, Watson2019, PaperI, PaperII, Perego2022, Vieira2023, Vieira2024}.

\begin{figure}
    \centering
    \includegraphics[width=\linewidth]{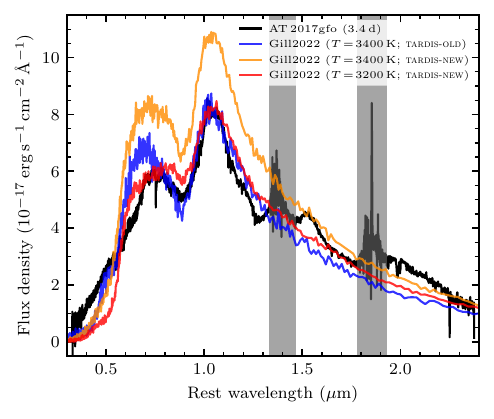}
    \caption{
        Comparison of the best-fitting \tardis\ model from \protect\cite{PaperI} (blue) with the observed 3.4\,d \xsh\ spectrum of \gfo\ (black). Regions of strong telluric absorption are shaded. We also show the resultant spectrum obtained by re-generating this best-fitting model with the updated \tardis\ code and its corrected relativistic treatment (orange), to illustrate the impact this correction has on the synthetic spectrum. Finally, we present the same model, but with $T$ re-scaled, to approximately account for this correction ($T = 3200$\,K, versus the original $T = 3400$\,K; red). We find that this re-scaled model closely resembles both the best-fitting model presented by \protect\cite{PaperI}, and the observed data.
    }
    \label{FIG: Demoing rescaled T model}
\end{figure}

Here we extend the work of \cite{PaperI}, with vastly improved atomic data inputs (as outlined in Section~\ref{SEC: Atomic data}). As in \cite{PaperI}, we reduce the parameter space we explore with our modelling by restricting ourselves to a limited set of composition profiles. In addition to consulting a Solar \rpro\ composition abundance (\eg, those presented by \citealt{Goriely1999} and \citealt{Prantzos2020}), we allow ourselves to also utilise the 13 abundance profiles presented by \cite{PaperI}, which were extracted from a nucleosynthetic post-processing calculation based on a realistic hydrodynamical simulation of a binary neutron star merger \citep[as presented by][]{Goriely2011, Goriely2013, Goriely2015, Bauswein2013}. Despite this apparent restriction, this limited set of `realistic' composition profiles successfully reproduced the $2.4 - 7.4$~day \xsh\ data of \gfo\ (specifically, \citealt{PaperI} were able to reproduce the observations invoking the \AngIII\ composition profile). For full details of model setup and approach, see \cite{PaperI}; here we only outline updates and differences to this approach, for brevity.

As in \cite{PaperI}, we utilise the \texttt{LTE} treatment for ionisation, \texttt{dilute-LTE} treatment for excitation, and \texttt{macroatom} line treatment to account for fluorescence effects -- an important effect for spectral feature formation in KNe. These quasi-LTE approximations within our \tardis\ simulations should be reasonable for modelling KN spectra at early times (see \eg, \citealt{Pognan2022_nlte}; although see \citealt{Brethauer2026} for a recent counterexample).

Since the work of \cite{PaperI}, the \tardis\ code has been updated to correct the special relativistic treatment of \cite{Vogl2019_TARDIS}.\footnote{\url{https://github.com/tardis-sn/tardis/pull/2159}} Specifically, a correction to the photon packet initialisation properties was required, as, at the beginning of a simulation, packets were being initialised at the inner boundary with properties sampled from a single-temperature blackbody function with an incorrect temperature; it was a factor ($2 \beta + 1)^{1/4}$ lower than expected (where $\beta = \frac{v}{c}$). Thus, re-running one of the `best-fitting' models presented by \cite{PaperI} with an updated version of the code returns a synthetic spectrum that is too luminous to match the observed data, as demonstrated in Figure~\ref{FIG: Demoing rescaled T model}.\footnote{We note that this is also the source of a similar discrepancy between model and observation in the work of \cite{Mulholland2024_Sr_Y}.}

A first-order fix to update the best-fitting \tardis\ models of \cite{PaperI} is to re-scale the input $T$ values by this correction factor. New \tardis\ models can be obtained that closely match those of \cite{PaperI} from this approach. In Figure~\ref{FIG: Demoing rescaled T model}, we present a model with $T$ re-scaled from $3400 \rightarrow 3200$\,K (given $v_{\rm inner} = 0.15$\,c, and thus $T_{\rm new} = T_{\rm old} / (2 \beta + 1)^{1/4} \simeq 3200$\,K). However, this updated inner boundary initialisation temperature has knock-on effects on the plasma properties (specifically, on ionisation and excitation). Thus, one should formally explore the full effects of this updated special relativistic treatment on the entire sequence of \gfo\ models presented by \cite{PaperI}. For now, we limit ourselves to re-analysing the 3.4\,d \xsh\ spectrum of \gfo\ \citep{Pian2017}, as the focus of this work is not on updates to the \tardis\ code base, but instead on the impact of updated input atomic data (specifically the calibrated line transition information of \citealt{Flors2026}). We utilise the flux-calibrated, de-reddened and de-redshifted spectral reduction from the ENGRAVE public data release,\footnote{\url{www.engrave-eso.org/AT2017gfo-Data-Release}} and convert the observational data to luminosity using $D_\textsc{l} = 40.4$\,Mpc \citep[][]{Hjorth2017}.

\section{Results \& Discussion} \label{SEC: Results}

\begin{figure}
    \centering
    \subfigure{\includegraphics[width=\linewidth]{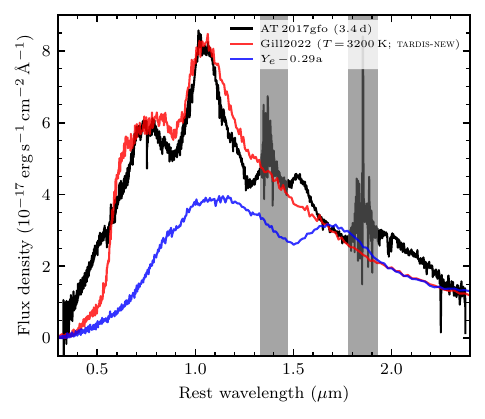}}\vspace{-1.5em}
    \subfigure{\includegraphics[width=\linewidth]{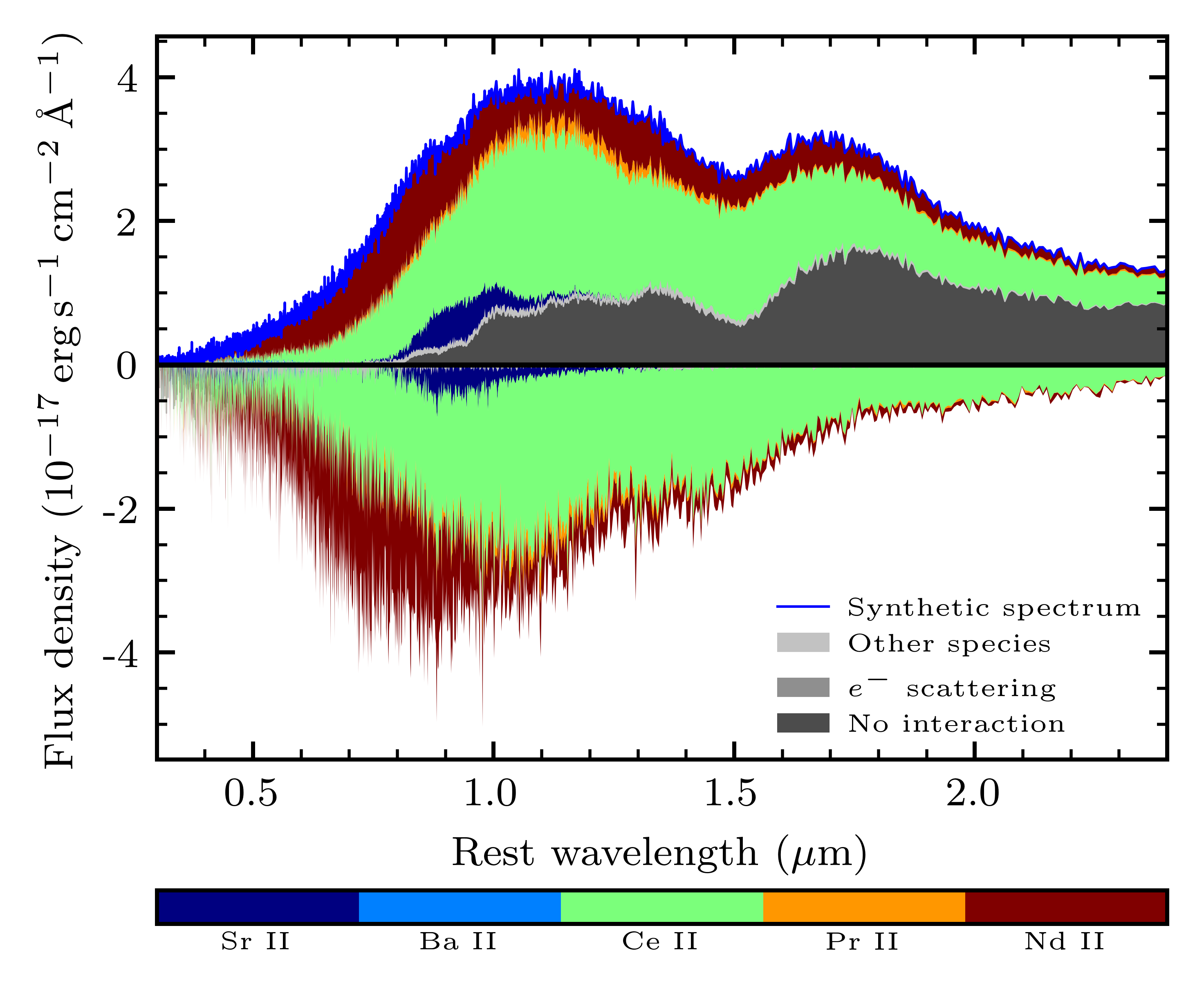}}
    \caption{
        \textit{Top:} Updated best-fitting models compared with the 3.4\,d \xsh\ spectrum of \gfo\ (black). Regions of strong telluric absorption in the observations are shaded. The re-scaled $T = 3200$\,K model that closely matches that of \protect\cite{PaperI} is again shown (red). The same model generated with our updated atomic data set is plotted for comparison (blue). Note how the entire SED has changed shape, due to the significantly increased lanthanide opacity.
        \textit{Bottom:} Model decomposition plot to illustrate the strongest contributions to the new model SED. Contributions from specific species are colour-coded; absorption is represented by shaded regions below Flux = 0, while emission is represented by shaded regions above Flux~=~0.
    }
    \label{FIG: 3.4d old vs new best-fitting model + SDEC}
\end{figure}

We begin our analysis by exploring a like-for-like comparison between the two atomic data sets (\citealt{PaperI} versus this work). In Figure~\ref{FIG: 3.4d old vs new best-fitting model + SDEC}, we compare the updated best-fitting \tardis\ model\footnote{Note that here, and throughout the rest of this manuscript, when we refer to the best-fitting model of \cite{PaperI}, we are referring to the synthetic spectrum presented in Figure~\ref{FIG: Demoing rescaled T model} which has been generated with a re-scaled $T$ to account for corrections to the \tardis\ code (dubbed \mbox{`Gill2022 ($T = 3200$\,K; \tardis-\textsc{new}'}). This allows us to remove any discrepancy between old and new model fits that are due to updates to the code base; \ie, any changes between this model and our other presented models will be due to updates to our input model properties only.} for the 3.4\,d \xsh\ spectrum of \gfo\ to an identical model with our updated compilation of atomic data. The only difference between these two models is the input atomic data.

The spectral energy distribution (SED) of the new model is completely altered, and now no longer matches the observed spectrum. Inspecting the different model contributions (see Figure~\ref{FIG: 3.4d old vs new best-fitting model + SDEC}), we see that this change is predominantly due to substantial line blanketing by the lanthanide species Ce\,\II\ and Nd\,\II. The old data set, which contained \dream\ data for the lanthanides, has now been replaced by \gsi\ data. In addition to a dramatic increase in the total number of transitions (see Table~\ref{TAB: GSI_vs_DREAM}), this new atomic data extends to much longer wavelengths, allowing substantial deviation from the inner boundary continuum in the near-infrared compared with the previous model (see Figure~8 in \citealt{PaperI}). As a specific example, Ce\,\II\ (the ion that dominates our line-forming region; see Figure~\ref{FIG: 3.4d old vs new best-fitting model + SDEC}) has 12,717 lines in the \dream\ data set, whereas the \gsi\ data contains 338,476 lines (a $27 \times$ increase; Table~\ref{TAB: GSI_vs_DREAM}). While there is of course some variation in the oscillator strengths for common lines in \dream\ and \gsi\ \citep{Flors2026}, which may result in a small subset of lines contributing more (or less) to the opacity in this new model, this effect is vastly subdominant to the effect of increasing the sheer number of available transitions in the model.

The inclusion of these new, significantly improved, atomic data immediately reveals the true impact of lanthanide elements in KN ejecta. We can see from Figure~\ref{FIG: 3.4d old vs new best-fitting model + SDEC} that the lanthanide contribution has been substantially underestimated by all previous works that have employed \tardis\ for modelling the spectra of \gfo, including \cite{PaperI}. This is a result of basing mass fraction estimates on vastly incomplete line transition information. Given the substantial impact of the lanthanides, we need to vastly reduce their total model contribution to better match the observed data. We can accomplish this by (i)~lowering the density of the model ejecta, thereby reducing the amount of lanthanide material in the model, or (ii)~exploring other compositions with lower lanthanide mass fractions.

For (i), we explore a wide parameter space of input model parameters (\eg, by varying $v_{\rm inner}$, $T$, $\rho_0$, $\Gamma$),\footnote{We invoke a power-law density structure with normalisation parameters, $t_0 = 2$\,d and $v_0 = 14000$\,km\,s$^{-1}$ \citep[see][]{PaperI}.} and find that updates to these model parameters alone are not sufficient; for example, while it is true that reducing the density in the model reduces the overly strong impact of lanthanides on the emergent spectrum, it also reduces the amount of Sr\,\II, thereby destroying model agreement with the Sr\,\II\ P-Cygni feature. For (ii), we find that none of the 13 composition profiles presented by \cite{PaperI} -- or a Solar \rpro\ composition -- can be invoked to reproduce the data. All models fail to reproduce the flux suppression at blue wavelengths ($\lesssim 0.7$\,\micron), while also producing a prominent P-Cygni feature to match the data between $\sim 0.7 - 1.2$\,\micron. This is a result of all composition profiles containing incompatible composition ratios of \mbox{Sr / lanthanides}. This statement alone contains some interesting insight; perhaps this indicates that the previously published composition profiles are not representative of characteristic kilonova ejecta, or at least it may indicate that they are incompatible with this specific event. Further quantification of this, with parameter-space expansion including other proposed kilonova ejecta compositions, should be undertaken.

Given we are unable to reproduce the data, we allow ourselves to deviate from these fixed composition profiles, to explore what type of abundance profile can reproduce the data. Fundamentally, to match the data, we require two things -- a source of flux suppression at blue ($\lesssim 0.7$\,\micron) wavelengths, and enough Sr\,\II\ to replicate the prominent $\sim 0.7 - 1.2$\,\micron\ P-Cygni profile. Given the lanthanide species appear to be the cause of the largest discrepancy with the data, we opt to alter the best-fitting \AngIII\ composition profile from \cite{PaperI} by re-scaling the lanthanide mass fractions.

\begin{figure}
    \centering
    \subfigure{\includegraphics[width=\linewidth]{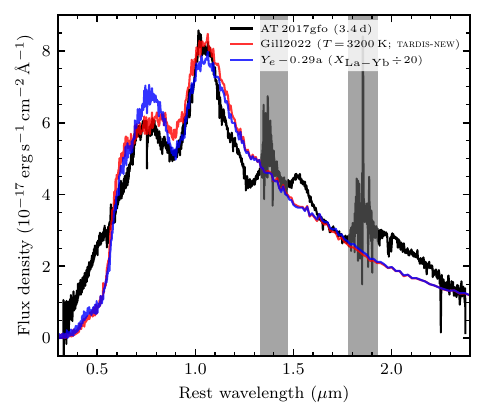}}\vspace{-1.5em}
    \subfigure{\includegraphics[width=\linewidth]{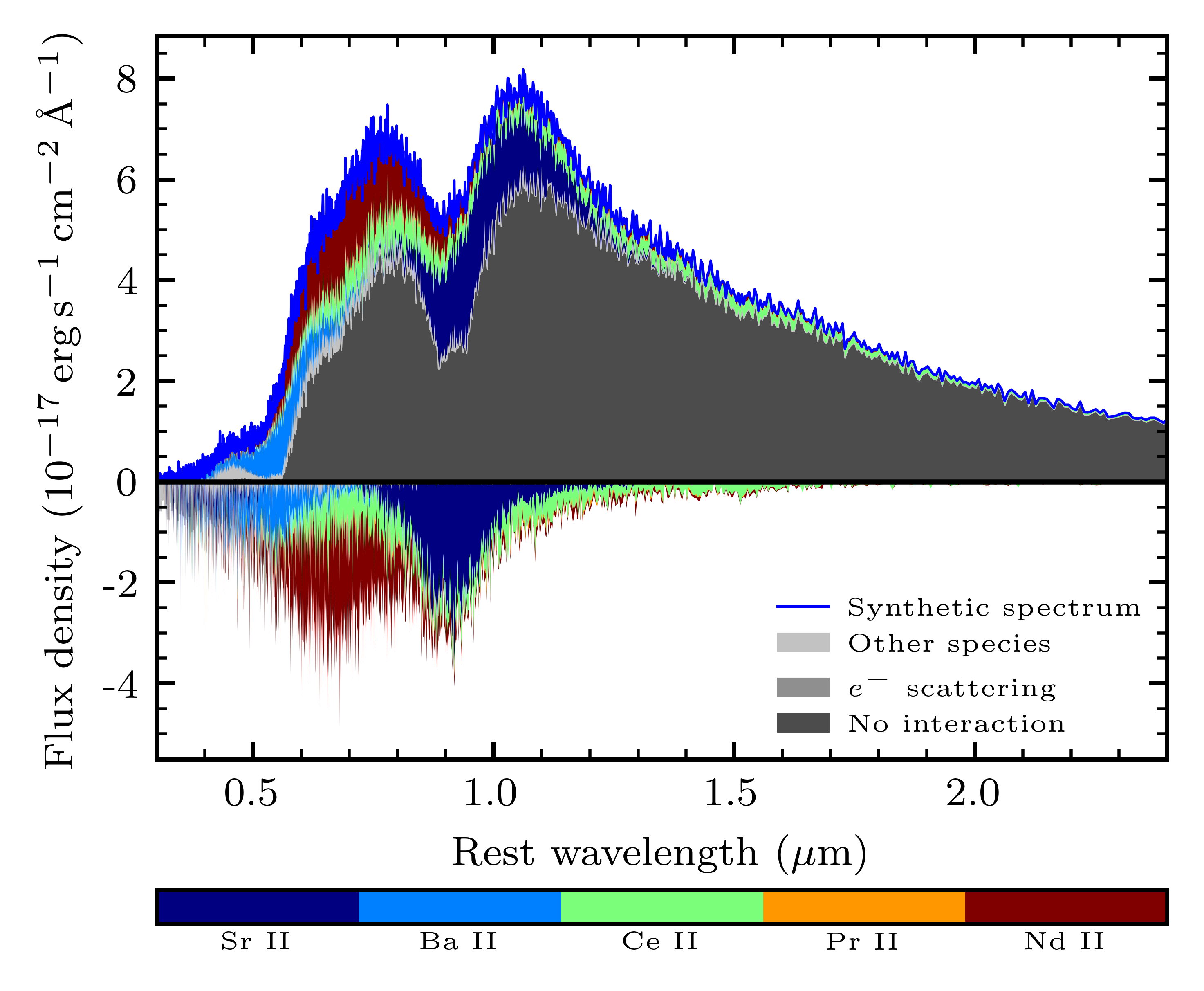}}
    \caption{
        \textit{Top:} Comparison between our new best-fitting model (blue), our old best-fitting model (red), and the 3.4\,d \xsh\ spectrum of \gfo\ (black). Regions of strong telluric absorption are again shaded. This best-fitting model was obtained with a modified version of the \AngIII\ composition profile.
        \textit{Bottom:} Model decomposition plot for our new best-fitting model.
    }
    \label{FIG: 3.4d new best-fitting model + SDEC}
\end{figure}

From this approach, we find that we can recover a much more reasonable fit to the data; specifically, by invoking a $20 \times$ reduction in the lanthanide ($Z = 57 - 70$) mass fraction.\footnote{Modelling of the $2.4 - 5.4$\,d X-shooter spectral sequence reveals that this modified composition profile can also reproduce those data; additionally, we find that a different composition is still required to model the earliest 1.4\,d X-shooter spectrum \citep[corroborating][]{PaperI}.} The input \tardis\ model properties of this updated best-fitting model are summarised in Table~\ref{TAB: TARDIS model properties}, while the model spectrum is compared to the 3.4\,d \xsh\ spectrum of \gfo\ in Figure~\ref{FIG: 3.4d new best-fitting model + SDEC}. This model more closely resembles the SED of the observed spectrum for $\lambda \geq 0.6$\,\micron, but contains prominent absorption at shorter wavelengths, at odds with the observed data (but mirroring the previous best-fitting \tardis\ model). This discrepancy is due to Ba\,\II. Ba\,\II\ appears to be able to create a prominent feature in the model if there is not some strong blanketing from other, heavier, \rpro\ species. Ba is a Group~2 element, and thus Ba\,\II\ is a homologue to both Ca\,\II\ and Sr\,\II; all three of these species have simple energy level structures, resulting in relatively few, but intrinsically very strong, permitted transitions. Ca\,\II\ and Sr\,\II\ have already been shown to produce strong imprints on kilonova spectra \citep{Domoto2021}, even while having modest abundances, and Ba\,\II\ is no different (see also \citealt{Domoto2022}).

\input{Tables/TARDIS_model_properties}

\begin{figure}
    \centering
    \includegraphics[width=\linewidth]{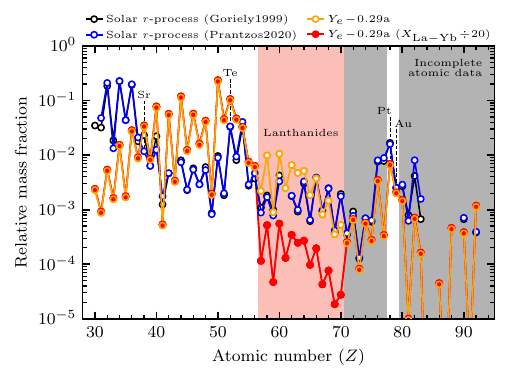}
    \caption{
        Comparison between the Solar \rpro\ distributions of \protect\cite{Goriely1999} (black) and \protect\cite{Prantzos2020} (blue), the \AngIII\ composition profile of \protect\cite{PaperI} (orange), and our re-scaled composition profile invoked in this work (red). All distributions have been re-scaled such that relative mass fractions for $Z = 30 - 92$ sum to unity. Elements of interest have been labelled, and shaded regions highlight the lanthanide elements (pink) and elements with very incomplete atomic data (grey).
    }
    \label{FIG: Composition profiles}
\end{figure}

\cite{PaperI} claim that the lanthanide mass fraction in the line-forming region of the ejecta $> 2$~days is $X_\textsc{ln} \simeq 0.05^{+0.05}_{-0.02}$. Here we find that this is a significant overestimate. Instead we find that, to reproduce the 3.4\,d \xsh\ spectrum of \gfo, we require $X_{\textsc{ln}} \approx 2.5 \times 10^{-3}$. Our new best-fitting composition profile is presented in Figure~\ref{FIG: Composition profiles}, where we compare it to both the Solar \rpro\ compositions of \cite{Goriely1999} and \cite{Prantzos2020}, as well as the unscaled \AngIII\ composition profile. The most striking result from this new fit to the observations (and evidenced most clearly in Figure~\ref{FIG: Composition profiles}) is that the new inferred abundance of the lanthanide species is significantly lower than the Solar \rpro\ distribution. Further study is needed to quantify the (somewhat arbitrary) composition profile we invoke here; but it, at least initially, appears to indicate that \gfo-like events may struggle to reproduce the heavy \rpro\ elements to the degree needed to match the Solar \rpro\ distribution. However, we caution that our modelling is only sensitive to the ejecta within the line-forming region; the dense inner regions of ejecta may contain heavier \rpro\ species, potentially resolving this apparent conflict. Our finding of a sub-Solar \rpro\ composition is in line with the compilation of $X_{\textsc{ln}}$ values presented by \cite{ji2019} ($X_{\textsc{ln}} = 2 \times 10^{-3} - 2 \times 10^{-2}$).

One point to note with regards to our new best-fitting composition profile, presented in Figure~\ref{FIG: Composition profiles}, is that our model possesses abundances for many trans-lanthanide (\ie, $Z \geq 71$) species that are higher than the lanthanide abundances. We are not claiming that our model necessitates this top-heavy composition profile, consisting of a high mass fraction of very heavy \rpro\ species. Instead, our modelling inferences for species with $Z = 71 - 92$ (excluding $_{78}$Pt and $_{79}$Au) are hampered by the extremely sparse atomic data (from \kuruczgfall) for these species. While we have very complete atomic data from \qub\ for Pt and Au \citep[][]{Gillanders2021, McCann2022}, these species have already been shown to only weakly contribute to photospheric-phase KN spectra \citep[][]{Gillanders2021}. Thus, our modelling is unable to probe abundances for species heavier than the lanthanides, and so the mass fractions for $Z = 71 - 92$, presented in Figure~\ref{FIG: Composition profiles}, should not be over-interpreted.

\section{Conclusions} \label{SEC: Conclusions}

In this manuscript, we have revisited previously published spectral modelling analysis of \gfo, with an emphasis on exploiting the newly published, and significantly more complete, line transition information for the lanthanide elements from \cite{Flors2026}. From our spectral modelling and analysis, we derive the following main conclusions:

\begin{itemize}

    \item The previously inferred effects of lanthanide material on the emergent SED of the KN models presented by \cite{PaperI} were significantly underestimated. The updated line transition information utilised here \citep[\gsi;][]{Flors2026} has shown that all previously utilised data sets were vastly incomplete, and that the number of transitions that were present in the previous data were not enough to quantitatively constrain lanthanide contribution.
    
    \item It is no longer possible to reproduce the 3.4\,d \xsh\ spectrum of \gfo\ with one of the `realistic' KN composition profiles presented by \cite{PaperI}. Whether this implies \gfo\ possessed an atypical composition, or that these profiles are not representative of the ejecta of KNe (or specifically \gfo) remains an open question.

    \item We can reproduce the data with a modified version of these KN composition profiles (see Figure~\ref{FIG: Composition profiles}). Our new best-fitting model suggests that the mass fraction of lanthanides in the line-forming region of \gfo\ at 3.4~days is $X_{\textsc{ln}} \approx 2.5 \times 10^{-3}$, $20 \times$ lower than that reported by \cite{PaperI} ($X_{\textsc{ln}} \simeq 0.05^{+0.05}_{-0.02}$). Further modelling works should be undertaken to determine whether this new lanthanide mass fraction estimate is compatible with the other epochs of observed spectra of \gfo, and/or whether a more physically motivated composition profile can satisfactorily reproduce the observed data.
    
    \item We find that to match the data, we need to invoke a significantly sub-Solar abundance for the lanthanides. KNe likely possess stratified compositions as a result of their multiple distinct ejection mechanisms; thus, we cannot constrain the bulk ejecta composition (and total ejected lanthanide mass) from this analysis. The inference of a sub-Solar lanthanide abundance in the line-forming region of \gfo, while intriguing, does not necessarily mean that \gfo-like KNe are incapable of producing the Solar \rpro\ distribution; further study should be undertaken to explore this more fully, in an effort to quantify lanthanide production in \gfo.

\end{itemize}

This study demonstrates that more complete, and crucially, accurately calibrated, line transition information is vital for spectroscopic studies such as that presented here. The recently published \gsi\ data from \cite{Flors2026} provides a substantial step forward for this, and other similar, studies. Accurately calibrated data now exist for a substantial number of relevant species across the first, second, and third \rpro\ peaks, in addition to the lanthanides. Further atomic data advances in the coming years will enable iterative improvement towards understanding the composition of kilonovae, thus ultimately leading to an improved understanding of their role as (one of) the site(s) of \rpro\ nucleosynthesis.

\section*{Acknowledgements}

We thank the referees for a detailed and constructive review. We thank Stuart~A.~Sim, Stephen~J.~Smartt and Leo~P.~Mulholland for stimulating discussions and constructive comments surrounding this project. AF acknowledges support by the European Research Council (ERC) under the European Union's Horizon 2020 research and innovation programme (ERC Advanced Grant KILONOVA No.~885281), the Deutsche Forschungsgemeinschaft (DFG, German Research Foundation) - Project-ID 279384907 - SFB 1245, and MA 4248/3-1. RFS acknowledges the support from National funding by FCT (Portugal), through the individual research grant 2022.10009 and through project funding \href{https://doi.org/10.54499/2023.14470.PEX}{2023.14470.PEX} ``Spectral Analysis and Radiative Data for Elemental Kilonovae Identification (SPARKLE)''. This research made use of \tardis, a community-developed software package for spectral synthesis in supernovae. The development of \tardis\ received support from GitHub, the Google Summer of Code initiative, and from ESA's Summer of Code in Space program. \tardis\ is a fiscally sponsored project of NumFOCUS. \tardis\ makes extensive use of \texttt{astropy} and \texttt{radioactivedecay}. We are grateful for use of the computing resources from the Northern Ireland High Performance Computing (NI-HPC) service funded by EPSRC (EP/T022175). We made use of the flux-calibrated \xsh\ spectra publicly available through ENGRAVE, which are based on observations collected at the European Southern Observatory (ESO), available through the ESO Science Archive Facility.

\section*{Data Availability}

The observational data presented are publicly available. All information needed to reproduce our modelling has been outlined in full.


\bibliographystyle{mnras}
\bibliography{References}

\bsp 
\label{lastpage}
\end{document}

%% file: Packages.tex
\usepackage{caption}

\usepackage{tikz}
\usepackage{orcidlink}
\usepackage{comment}

\usepackage{graphicx}

\usepackage{multirow}
\usepackage{xcolor}
\usepackage{manyfoot}
\usepackage{booktabs}
\usepackage{changepage}
\usepackage[percent]{overpic}
\usepackage{xfrac}

\usepackage[normalem]{ulem}

\usepackage{booktabs,threeparttable}

\usepackage{amsmath,bm}

\usepackage{subfigure}

%% file: Commands.tex
\newcommand{\tardis}{\mbox{\textsc{tardis}}}
\newcommand{\carsus}{\mbox{\textsc{carsus}}}

\newcommand{\chianti}{\mbox{\textsc{chianti}}}
\newcommand{\kuruczgfall}{\mbox{\textsc{gfall}}}
\newcommand{\kuruczatoms}{\mbox{\textsc{atoms}}}
\newcommand{\dream}{\mbox{\textsc{dream}}}
\newcommand{\gsi}{\mbox{\textsc{gsi}}}
\newcommand{\qub}{\mbox{\textsc{qub}}}

\newcommand{\I}{{\sc i}}
\newcommand{\II}{{\sc ii}}
\newcommand{\III}{{\sc iii}}
\newcommand{\IV}{{\sc iv}}

\newcommand{\eg}{\mbox{e.g.}}
\newcommand{\ie}{\mbox{i.e.}}

\newcommand{\rpro}{\mbox{$r$-process}}

\newcommand{\xsh}{\mbox{X-shooter}}

\newcommand{\gfo}{\mbox{AT\,2017gfo}}

\newcommand{\ditto}{\texttt{"}}

\newcommand{\Ye}{$Y_e$}
\newcommand{\AngIII}{\Ye$- 0.29$a}

%% file: Tables/AtomicDataSources.tex
\begin{table}
    \centering
    \caption{
        Summary of the different atomic data sources incorporated into our \tardis\ atomic data file, compared to that of \protect\cite{PaperI}.
    }
    \begin{threeparttable}
        \centering
        \begin{tabular}{lcc}
            \toprule
            Species                                                 &This work        &\protect{\cite{PaperI}}                                          \\
            \midrule
    
            $_1$H, $_2$He                                           &\chianti         &\chianti                                                         \\
            $_{3}$Li -- $_{37}$Rb                                   &\kuruczgfall     &\kuruczgfall                                                     \\
            $_{38}$Sr -- $_{40}$Zr                                  &\kuruczatoms     &\kuruczatoms                                                     \\
            $_{41}$Nb -- $_{56}$Ba                                  &\kuruczgfall     &\kuruczgfall                                                     \\
            \bm{$_{57}$}\textbf{La} -- \bm{$_{70}$}\textbf{Yb}      &\textbf{\gsi}    &\textbf{\dream\ {\scriptsize \&} \kuruczgfall\tnote{$\star$}}    \\
            $_{71}$Lu                                               &\dream           &\dream                                                           \\
            $_{72}$Hf -- $_{77}$Ir                                  &\kuruczgfall     &\kuruczgfall                                                     \\
            $_{78}$Pt, $_{79}$Au                                    &\qub             &\qub                                                             \\
            $_{80}$Hg -- $_{92}$U                                   &\kuruczgfall     &\kuruczgfall                                                     \\

            \bottomrule
        \end{tabular}
    \begin{tablenotes}
        \small
        \item[$\star$] \dream\ does not contain data for La\,\II, Eu\,\II, Gd\,\II, Tb\,\II, Dy\,\II\ and Ho\,\II; \cite{PaperI} default to \kuruczgfall\ for these ions.
    \end{tablenotes}
    \end{threeparttable}
    \label{TAB: Atomic data}
\end{table}

%% file: Tables/GSI_vs_DREAM.tex
\begin{table}
    \centering
    \caption{
        Number of lines present in the \gsi\ and \protect\cite{PaperI} (\dream\ {\scriptsize \&} \kuruczgfall) data sets that satisfy $\log(gf) \geq -3$.
    }
    \begin{threeparttable}
        \centering
        \begin{tabular}{rlrrcrr}
            \toprule
            Element         &Ion        &\multicolumn{5}{c}{No.~of~lines}                                                       \\
            \cmidrule{3-7}
                            &           &\multicolumn{2}{c}{\gsi}               &        &\multicolumn{2}{c}{\cite{PaperI}}     \\
            \cmidrule{3-4}
            \cmidrule{6-7}
                            &           &All $\lambda$      &$> 1$\,\micron     &        &All $\lambda$     &$> 1$\,\micron     \\
            \midrule

            $_{57}$La       &\II        &15,961             &5,118              &        &273               &1                  \\
            \ditto          &\III       &205                &52                 &        &129               &0                  \\
            $_{58}$Ce       &\II        &338,476            &97,689             &        &12,717            &2                  \\
            \ditto          &\III       &6,230              &644                &        &2,391             &0                  \\
            $_{59}$Pr       &\II        &424,350            &43,222             &        &144               &0                  \\
            \ditto          &\III       &55,767             &2,746              &        &12,986            &0                  \\
            $_{60}$Nd       &\II        &2,357,380          &307,542            &        &106               &0                  \\
            \ditto          &\III       &514,365            &35,636             &        &43                &0                  \\
            $_{61}$Pm       &\II        &415,689            &34,646             &        &0 		        &0                  \\
            \ditto          &\III       &538,011            &14,900             &        &0 		        &0                  \\
            $_{62}$Sm       &\II        &578,557            &19,981             &        &162               &0                  \\
            \ditto          &\III       &1,134,068          &36,465             &        &73                &0                  \\
            $_{63}$Eu       &\II        &385,059            &48,152             &        &159               &6                  \\
            \ditto          &\III       &734,893            &24,168             &        &788               &0                  \\
            $_{64}$Gd       &\II        &1,614,181          &77,469             &        &779               &0                  \\
            \ditto          &\III       &516,465            &2,759              &        &44                &0                  \\
            $_{65}$Tb       &\II        &3,147,305          &50,576             &        &104               &0                  \\
            \ditto          &\III       &856,585            &14,670             &        &793               &0                  \\
            $_{66}$Dy       &\II        &2,435,744          &49,310             &        &790               &0                  \\
            \ditto          &\III       &207,615            &128                &        &1,152             &0                  \\
            $_{67}$Ho       &\II        &446,994            &13,706             &        &12                &0                  \\
            \ditto          &\III       &285,764            &12,427             &        &1,068             &0                  \\
            $_{68}$Er       &\II        &657,642            &22,205             &        &19                &0                  \\
            \ditto          &\III       &120,180            &7,391              &        &1,065             &0                  \\
            $_{69}$Tm       &\II        &92,279             &3,659              &        &6,317             &0                  \\
            \ditto          &\III       &9,088              &605                &        &1,177             &0                  \\
            $_{70}$Yb       &\II        &7,486              &606                &        &5,112             &667                \\
            \ditto          &\III       &903                &32                 &        &230               &0                  \\

            \midrule

            \multicolumn{2}{c}{Total}   &17,897,242         &926,504            &        &48,633            &676                \\

            \bottomrule
        \end{tabular}
    \end{threeparttable}
    \label{TAB: GSI_vs_DREAM}
\end{table}

%% file: Tables/TARDIS_model_properties.tex
\begin{table}
    \centering
    \caption{
        Input properties invoked for our updated best-fitting \tardis\ model, and those of the best-fitting model of \protect\cite{PaperI}.
    }
    \begin{threeparttable}
        \centering
    \begin{tabular}{lcc}
        \toprule
                                    &This work                  &\protect{\cite{PaperI}}        \\
        \midrule
        
        $t_{\rm exp}$ (d)           &3.4                        &3.4                            \\
        $v_{\rm inner}$ (c)         &0.15                       &0.15                           \\
        $v_{\rm outer}$ (c)         &0.35                       &0.35                           \\
        $\rho_0$ (g\,cm$^{-3}$)     &$4 \times 10^{-15}$        &$4 \times 10^{-15}$            \\
        $\Gamma$                    &$-3$                       &$-3$                           \\
        Composition profile         &\AngIII\tnote{$\star$}     &\AngIII                        \\
        $T$ (K)                     &3200                       &3400\tnote{$\square$}          \\
        
        \bottomrule
    \end{tabular}
    \begin{tablenotes}
        \small
        \item[$\star$] This model utilises a modified version of the \AngIII\ composition profile (the relative mass fractions of $_{57}$La -- $_{70}$Yb have been rescaled; see Figure~\ref{FIG: Composition profiles}).
        \item[$\square$] This temperature is the best-fitting value derived from an older version of the \tardis\ code (see Section~\ref{SEC: Model setup}).
    \end{tablenotes}
    \end{threeparttable}
    \label{TAB: TARDIS model properties}
\end{table}